\documentclass[11pt]{article}

\usepackage{float}
\floatstyle{boxed}
\restylefloat{figure}

\usepackage{amsmath,amssymb,amsthm,amscd,latexsym}
\usepackage[toc,title,titletoc]{appendix}
\usepackage{cite}
\usepackage{graphicx,epsfig}
\usepackage{breakurl}

\RequirePackage{color}

\allowdisplaybreaks[1]

\def\J2{J_2}

\def\virg#1{``#1"}

\def\eqi{\begin{equation}}
\def\eqf{\end{equation}}
\def\eqia{\begin{eqnarray}}
\def\eqfa{\end{eqnarray}}

\begin{document}

\title{A new type of misconduct in the field of the Physical Sciences: the case of the pseudonyms used by I. Ciufolini to anonymously criticize other people's works on arXiv}

\author{L. Iorio\\ Ministero dell'Istruzione, dell'Universit$\grave{\textrm{a}}$ e della Ricerca (M.I.U.R.)-Istruzione \\ Viale Unit$\grave{\textrm{a}}$ di Italia 68, 70125, Bari (BA), Italy\\ email: lorenzo.iorio@libero.it}

\maketitle


%
Dear Sir,
in this Letter two cases of fraudulent scientific conduct in the field of the Physical Sciences that occurred in 2007 on the arXiv pre-print repository maintained by the Cornell University are reported.

They pertain the so-called gravitomagnetic frame-dragging, or Lense-Thirring effect. It is a prediction of the General Theory of Relativity (GTR) about the behaviour of gyroscopes and satellites moving in the gravitational field of a rotating body such as, e.g., the Earth. It has been the subject of intense experimental scrutiny in recent years.
Perhaps, the most famous experiment aimed to testing frame-dragging is the spacecraft-based Gravity Probe B (GP-B) mission\footnote{See \textrm{http://einstein.stanford.edu/} on the Internet}. It was launched in 2004, and its Principal Investigator was F. Everitt. The results of its data analysis were finally announced in May 2011, reporting a successful measurement of the relativistic tiny shifts of the axes of four gyroscopes carried onboard  within a claimed  margin of uncertainty of 19$\%$.
A direct competitor of GP-B is the ongoing analysis of the laser-ranged observations of the geodetic satellites of the LAGEOS family performed by I. Ciufolini and coworkers. In October 2004, they reported a successful detection of the Lense-Thirring shift of the satellites' orbital planes with a claimed $10\%$ accuracy. Since then, such a level of uncertainty has been often questioned in several papers by L. Iorio and other researchers; it may be larger by a factor $2-3$ or, perhaps, even more.
In 2006, L. Iorio suggested that the Lense-Thirring effect may have been detected by the Mars Global Surveyor (MGS) spacecraft orbiting Mars with a few percent accuracy.
%
%

In March 2007, a certain G. Felici suddenly popped up on arXiv. He posted a pre-print \cite{2007gr.qc.....3020F}  criticizing the interpretation of the MGS data proposed by the author of this Letter. G. Felici, who has not published anything neither before nor after that date, claimed to be based in via Attilio Regolo 2, 20138, Milano, Italy, and he used the email gia.felici@yahoo.it. To date, his pre-print has not yet been published in any peer-reviewed journal.
In December 2013, the arXiv's moderators  added the following comment to the Felici's pre-print: \virg{This submission has been made by G. Felici, a pseudonym of Ignazio Ciufolini, who repeatedly submits inappropriate articles under pseudonyms, in violation of arXiv policies}.

In December 2007, a certain G. Forst, who never had published anything before,  appeared on arXiv. He posted a manuscript \cite{2007arXiv0712.3934F} criticizing certain aspects of the GP-B mission. His email, g.forst@yahoo.com, was not an academic one, and it could not be possible to retrieve the alleged Forst's institution, FGP Behrenstr. 1 10117 Berlin, on the Internet.
%
%
%
%
%
%
%
%
In January 2008, the arXiv's moderators retracted the Forst's pre-print with the following comment: \virg{This submission has been removed because `G.Forst' is a pseudonym of a physicist based in Italy who is unwilling to submit articles under his own name. This is in explicit violation of arXiv policies.
Roughly similar content, contrasting the relative merits of the LAGEOS and GP-B measurements of the frame-dragging effect, can be found in pp. 43-45 of: \cite{2007Natur.449...41C}}.
%
%
%
%
%
%
Nonetheless, I. Ciufolini was the sole author in the scientific community who frequently cited the Forst's pre-print in talks and presentations given at international meetings and institutions, pre-prints and peer-reviewed papers.
%
%
%
%
%
%
%
%
%
%
%
After 6 years, in  September 2013\footnote{Private communication by the arXiv's moderators to this author, 19 December 2013.}, the arXiv's moderators altered their original comment by writing: \virg{This submission has been removed because `G.Forst' is a pseudonym of Ignazio Ciufolini, who repeatedly submits inappropriate articles under pseudonyms. This is in explicit violation of arXiv policies. Roughly similar content, contrasting the relative merits of the LAGEOS and GP-B measurements of the frame-dragging effect, can be found in pp. 43-45 of: \cite{2007Natur.449...41C}}.
%
%
%
%
%
%
In May 2011, I. Ciufolini congratulated the GP-B team after they released their results by stating that: \virg{GP-B was a beautiful and challenging experiment}, as reported in an online blog run by Nature.
%
%
%
%
%
%
%
%
%
%

%
%
%
%
%
%
%

%
%
%
%
%

At present, the author of this Letter does not know why the arXiv's moderators disclosed the real identity of G. Felici and G. Forst after 6 years since their original submissions.

Full details and screenshots can be found on the Internet at http://gravityprobebpseudonyms.wordpress.com/ and at https://www.facebook.com/GPBframedraggingArXivCiufolini
\bibliography{ArXivbib}{}

\providecommand{\href}[2]{#2}\begingroup\raggedright\begin{thebibliography}{1}

\bibitem{2007gr.qc.....3020F}
G.~{Felici}, ``{The meaning of systematic errors, a comment to ''Reply to On
  the Systematic Errors in the Detection of the Lense-Thirring Effect with a
  Mars Orbiter'', by Lorenzo Iorio},'' {\em ArXiv General Relativity and
  Quantum Cosmology e-prints} (Mar., 2007) ,
  \href{http://arxiv.org/abs/gr-qc/0703020}{{\ttfamily gr-qc/0703020}}.

\bibitem{2007arXiv0712.3934F}
G.~{Forst}, ``{A critical analysis of the GP-B mission. I: on the impossibility
  of a reliable measurement of the gravitomagnetic precession of the GP-B
  gyroscopes},'' {\em ArXiv e-prints} (Dec., 2007) ,
  \href{http://arxiv.org/abs/0712.3934}{{\ttfamily arXiv:0712.3934 [gr-qc]}}.

\bibitem{2007Natur.449...41C}
I.~{Ciufolini}, ``{Dragging of inertial frames},''
  \href{http://dx.doi.org/10.1038/nature06071}{{\em Nature} {\bfseries 449}
  (Sept., 2007) 41--47}.

\end{thebibliography}\endgroup

\end{document}